# On the number of nucleoproteins in the assembly of coronaviruses: Consequences for COVID-19

**Vladimir R. Chechetkin[1*] and Vasily V. Lobzin[2]**

[1]Engelhardt Institute of Molecular Biology of Russian Academy of Sciences, Vavilov str., 32, Moscow, Russia

[2]School of Physics, University of Sydney, Sydney, NSW 2006, Australia

*Corresponding author. Vladimir R. Chechetkin

vladimir_chechet@mail.ru

**Abstract**

The multifunctional nucleoproteins play important role in the life cycle of coronaviruses. The assessment of their quantities is of general interest for the assembly of virions and medical applications. The proliferating nucleoproteins induce the related (auto)immune response and via binding to host RNA affect various regulation mechanisms. In this report we briefly summarize and comment the available experimental data on the subject concerned.

**Keywords:** coronaviruses; nucleoproteins; assembly of virions; antibodies; host RNA; COVID-19

At a state-of-the-art knowledge on the assembly of coronaviruses, there remains significant divergence between the estimates on the numbers of nucleoproteins (N) in the virions. A part of methods provides that the complete coverage of the ssRNA coronavirus genome of length about 30,000 nt by the nucleocapsid needs $0.7–2.2\times10^3$ N proteins per virion or that each N is associated with 14–40 nt of genomic RNA [1], whereas the others yield that each N is associated approximately with 7 nt or the number of N proteins per virion amounts to $4.3\times10^3$ [2, 3]. According to [2, 3], a turn of helical nucleocapsid for SARS-CoV comprises eight N proteins and is associated with about 56 nt. The nucleocapsid for SARS-CoV has an outer diameter of 16 nm, an inner diameter of 4 nm and a pitch of 14 nm [2, 3]. N proteins are evolutionary conserved and the similar parameters were determined for MHV nucleocapsid [4]. A helical curve with the parameters of ssRNA wound onto the inner nucleocapsid surface would have about 54–56 nt per turn [5].

The genomic ssRNA of coronaviruses is packaged within helical nucleocapsid. The transitional symmetry of the helical nucleocapsid of coronaviruses and weakly specific cooperative interactions between ssRNA and N proteins should lead to the natural selection of specific quasi-periodic assembly/packaging signals (here both terms concern the assembly of ribonucleocapsid) in the related genomic sequence. Such signals coordinated with the nucleocapsid helical structure were detected and reconstructed in the genomes of the coronaviruses by bioinformatic methods [5–7]. The pitch of detected periodicity was 51–57 nt that was close to ssRNA associated with the pitch of ribonucleocapsid helix obtained in [2, 3]. The periodicity of 54 nt was especially strongly pronounced in the genomes of the lineage B betacoronaviruses including the most pathogenic coronaviruses SARS-CoV and SARS-CoV-2. Quasi-periodicity for the genomes longer and shorter than that of SARS-CoV was biased to 57 nt and to 51 nt, respectively. This indicates that the total number of N proteins, $4.4\times10^3$, remains approximately conserved.

Forsythe et al. [8] studied the binding of N proteins with ssRNA of SARS-CoV-2 in solution containing 1000-nt fragments of ssRNA. They found that about 70 N proteins were bound to a 1000-nt fragment that corresponds to one N per 14 nt. The biomolecular N-RNA condensates provide a plausible model for N-RNA complexes in cellular cytosol after freeing encapsidated viral RNA from envelope. The solution with N-RNA complexes undergoes liquid-liquid phase separation (LLPS) by the phases enriched and depleted by N-RNA complexes [9–12].

A large part (a half or more according to [8]) of N proteins dissociates from ribonucleocapsid after freeing from envelope and should somehow diffuse and proliferate within host cells. The partial uncovering of RNA is needed for the translation of viral RNA by host cell ribosomes that is the first synthetic event in the replication cycle of coronaviruses. Such unbinding and diffusion of N proteins has direct consequences for COVID-19. A majority of available vaccines against COVID-19 contains attenuated fragments of spike (S) proteins as one of the main components. As should be expected in the light of the above arguments, the prevalent antibodies in the serum from vaccinated patients should be associated with S proteins, whereas the prevalent antibodies in the serum from unvaccinated patients should be associated with N proteins. This appears to be in complete agreement with the observations by Savvateeva et al. [13] who used microarray technique for the detection of antibodies. The antibodies to SARS-CoV-2 N proteins can bind to human proteins and reveal cross-reactivity to own human proteins [14]. Liu et al. [15] found that in the repertoires of autoantibodies to 91 autoantigens typical of classic autoimmune diseases, the autoantibodies related to La ribonucleoprotein domain family member 1 (LARP1), a N protein interaction partner of SARS-CoV-2, dominate in the cohort of COVID-19 patients and persist even after 6 months from the moment of illness.

The genes coding for the structural proteins of coronaviruses, spike (S), envelope (E), membrane (M), and nucleoprotein (N), are conservably ordered along genomes as S-E-M-N [1]. The expression of genes coding for the structural proteins in coronaviruses is performed via nested subgenomic mRNAs (sgmRNAs) [16–19]. The nested sgmRNAs serve for the translation of fragments containing the genes S-E-M-N, E-M-N, M-N, and N. Among the reasons explaining such discontinuous transcription [16–19], the arguments related to quite different number of copies for nonstructural proteins and for the structural proteins needed for the assembly of virions may also be relevant. The latter hypothesis implies the abundance of N proteins in comparison with the others.

The synthesis of sgmRNAs is governed by transcription regulation sequences (TRS) preceding the genes coding for the structural and accessory proteins with one TRS related to leader sequence being located at the 5'-end of the genome [20–22]. TRS are conserved within particular lineages of coronaviruses. The secondary RNA structure associated with TRS is also conserved within the lineages. The core motif for TRS in lineage B coronaviruses is ACGAAC surrounded by 1-2 variable characters. There are nine such motifs in the genome of SARS-CoV-2. The start positions of the motifs ACGAAC on the reference SARS-CoV-2 genome (GenBank accession: NC_045512.2) are as follows: 70 (related to the leader sequence); 21556 (precedes the gene coding for S protein); 25385 (precedes the gene coding for ORF3a protein); 26237 (precedes the gene coding for E protein); 26473 (precedes the gene coding for M protein); 27041 (located at the end of the gene coding for M protein and precedes the gene coding for ORF6 protein); 27388 (precedes the gene coding for ORF7a protein); 27888 (precedes the gene coding for ORF8 protein); and 28260 (precedes the gene coding for N protein). The combinatorics of sgmRNAs containing the genes coding for S, E, M, and N assessed by the positions of TRS would correspond to the ratios 1:3:4:8. Neuman et al. [23] estimated that the number of dimeric complexes $M_2$ is about 1100 $M_2$ (or 2200 M) per virion. This implies the ratio M:N equal approximately to 1:2 at the number of nucleoproteins equal to 4400 N. The estimates for the ratio of spike trimers $S_3$ to M were ranged from $1S_3$:16M to $1S_3$:25M [2]. The lower boundary S:M ≈ 1:5 is also not far from the ratio 1S:4M obtained by sgmRNA combinatorics. Bar-On et al. [24] collected the most typical numerical parameters reported to 2020 in the literature related to SARS-CoV-2. The typical numbers of copies per virion according to their data were 2000M:1000N:100$S_3$:100E. The reported value 1000 N contradicts even to the experiments by Forsythe et al. [8] that provide the lower boundary for the number of nucleoproteins in the ribonucleocapsid and is in strong disagreement with the cryo-EM analysis and conclusions by Chang et al. [2].

The conserved elements in the genome of SARS-CoV-2 were proposed to try as potential antiviral drug targets [25, 26]. We should note that TRS provide very promising targets for such purposes. The multiplicity of TRS, the unified core sequences and conserved local secondary structure provide additional advantages for therapeutic targeting. The corresponding drug would simultaneously prevent the infection by SARS-CoV and SARS-CoV-2 as well as would inhibit zoonotic transmission of related lineage B coronaviruses to human organisms. An attenuated fragments of enzyme subunit responsible for TRS recognition can be tried as an agent binding to TRS and blocking TRS instead of compounds, aptamers or antisense oligonucleotide therapeutics. The binding of a compound to such subunit would also inhibit the replication of coronaviruses.

The protein-coding genes occupy no more than 2% of the human genome, while 75–85% of the genome is transcribed [27]. The related RNA is involved into numerous regulatory mechanisms. The proliferating N proteins strongly binding to viral RNA can also bind to host RNAs and affect regulation [28]. Besides the binding of N proteins to host RNAs, it was shown that binding of SARS-CoV-2 N proteins to the host 14-3-3 protein in the cytoplasm can regulate nucleocytoplasmic N shuttling [29]. Earlier, it was proved that SARS-CoV N protein can activate an AP-1 pathway, which regulates many cellular processes, including cell proliferation, differentiation and apoptosis [30, 31]. The critical involvement of nucleoproteins in the replication of coronaviruses, in the regulation of expression of viral genes and in virion assembly is well-known (see, e.g., [1, 2, 32] and the further references therein) and is beyond the scope of this note. The general discussion of SARS-CoV-2–human protein contactome can be found in [33].

N proteins are evolutionary the most conserved among the structural proteins of coronaviruses [1]. Therefore, they can be efficiently used for detection of various coronavirus strains. The related detection techniques for N proteins are rapidly developing [34, 35]. The suggestions of using nucleoproteins as therapeutic targets are commonplace [36]. It was found that a compound known as PJ34 inhibits coronavirus replication and the RNA-binding activity of HCoV-OC43 N protein [37]. One of the thread guiding strategies to neutralize the functional activity of N proteins consists in binding negatively charged agents to positively charged regions of N proteins contacting with RNA. This can be achieved, in part, by heparin [38], chicoric acid [39] and phytoconstituents from natural nutrition [40]. Recently, Luan et al. [41] found that an antibiotic, ceftriaxone sodium, can block the binding of RNA to N-NTD and inhibit the formation of the RNP complex.

The examples above show that the abundance and proliferation of N proteins play important role in virus-host interactions and potential medical applications. To sum up, the proper quantitative assessment of the number of nucleoproteins is essential for the details of coronavirus assembly and for the assessment of pathogenic impact of nucleoproteins on the host cells. The earlier estimates [1, 24] should be reconsidered.